\documentclass[showkeys,superscriptaddress,floatfix,aps,10pt,prd]{revtex4-2}
\usepackage{graphicx,epstopdf}
\usepackage[caption=false]{subfig}
\usepackage{appendix}
\usepackage[T1]{fontenc}
\usepackage{lmodern}
\usepackage[dvipsnames,x11names]{xcolor}
\usepackage[colorlinks=true,linkcolor=ForestGreen,citecolor=ForestGreen,urlcolor=NavyBlue]{hyperref}
\usepackage[sort&compress]{natbib}
\usepackage{lipsum}
\usepackage{morefloats}
\usepackage[pdf]{pstricks}
\usepackage{amsmath}
\usepackage{amssymb}
\usepackage{amsfonts}
\usepackage{rotating}
\usepackage{cancel}
\usepackage{mathtools}
\usepackage{bbm}
\usepackage{dsfont}
\usepackage{bbold}
\usepackage{multirow}
\usepackage{ulem}
\usepackage{physics}
\usepackage{orcidlink}
\usepackage{colortbl}

\def\xslash{x\!\!\!\slash }

\begin{document}

\title{Exploring electromagnetic characteristics of the vector and axial-vector $B_c$ mesons}

\author{Ula\c{s}~\"{O}zdem\orcidlink{0000-0002-1907-2894}}%
\email[]{ulasozdem@aydin.edu.tr }
\affiliation{ Health Services Vocational School of Higher Education, Istanbul Aydin University, Sefakoy-Kucukcekmece, 34295 Istanbul, T\"{u}rkiye}

 
\begin{abstract}
The magnetic moments of the $B_c$ mesons provide significant insights into their inner structure and geometric shape. Furthermore, a comprehensive understanding of the electromagnetic characteristics of $B_c$ mesons is essential for advancing our knowledge of confinement and heavy flavor effects. In light of this, we proceed to extract the magnetic moments of the ground-state vector and axial-vector $B_c$ mesons through the medium of the QCD light-cone sum rules. The magnetic moments of the axial-vector and vector $B_c$ mesons are found to be $\mu_{B_c}= -0.47 \pm 0.07~\mu_N$, and $\mu_{B_c}= 0.15 \pm 0.02~\mu_N$, respectively. A comparison of our results for the vector $B_c$ meson with other theoretical predictions has revealed discrepancies between the various predictions, which could prove useful as a complementary tool for interpreting the vector $B_c$ meson. The current experimental data set is limited to a small number of observed states of beauty-charm mesons. However, theoretical studies can play a valuable role in elucidating their nature and guiding future experimental investigations.
\end{abstract}

\maketitle

\section{motivation}\label{motivation}

A fundamental issue in both theoretical and experimental hadron physics is the understanding of hadron structures and their transitions at the Fermi scale. 
Heavy quarkonium mesons and triply-heavy baryons are of significant importance in the study of the relations between perturbative and non-perturbative QCD, as well as in the comprehension of heavy quark dynamics, considering the lack of light quark effects. The bottom-charm quarkonium states, or $B_c$ mesons, are of particular interest due to their composition of heavy quarks from disparate flavors. The ground and excited $B_c$ mesons located below the $BD$,   $BD^*$, $B^*D$, and $B^* D^*$ thresholds are prevented from annihilating into gluons, resulting in greater stability than that observed in the related charmonium and bottomonium mesons. In contrast to the substantial body of research that has been conducted on the charmonium and bottomonium sectors, the bottom-charm meson sector remains relatively poorly explored. Indeed, only a small number of low-lying $B_c$ states have been identified in experimental studies to date. 

The experimental discovery of $B_c$ mesons, which constituted a significant advancement in the field of particle physics, commenced in 1998. On this date,  
the CDF Collaboration announced the discovery of a $B_c$ meson  by means of the $B^{\pm} \rightarrow J/\psi \ell^{\pm} \nu$ decay. The measured mass was $M_{B_c} =6.40 \pm 0.39 \pm 0.13$ GeV~\cite{CDF:1998ihx}. Following experimental research conducted by various collaborations, the presence of the $B_c$ mesons has been confirmed employing the analysis of different decay mechanisms, such as the $B_c \rightarrow J/\psi  \pi $, $B^+_c \rightarrow B^0_s \pi^+$, $B_c^+ \rightarrow J/\psi \pi^+  \pi^- \pi^+ $, and $ B_c \rightarrow J/\psi K^+ K^- \pi^+$ processes \cite{CDF:1998ihx, CDF:2005yjh, CDF:2007umr, D0:2008bqs, LHCb:2012ag, LHCb:2013vrl, LHCb:2013kwl, LHCb:2013hwj, LHCb:2013rud}. In 2014, the ATLAS Collaboration announced the discovery of a meson state that was compatible with the predicted $B_c(2S)$ state. The reported mass was $M_{B_c(2S)} = 6842\pm 9$ MeV~\cite{ATLAS:2014lga}. Furthermore, the LHCb and CMS Collaborations reported the excited $B_c (2^1S_0)$ and $B_c(2^3 S_1)$ states in the $B^+_c \pi^+\pi^-$ invariant mass spectrum. The determined masses were  $M_{B_c (2^1S_0)}= 6872.1 \pm 2.2$ MeV and $M_{B_c(2^3 S_1)}= 6841.2\pm1.5$ MeV~\cite{CMS:2019uhm, LHCb:2019bem}. Nevertheless, the Particle Data Group lists only two $B_c$ mesons, which are designated the $B_c(1S)$ and $B_c(2S)$ states~\cite{ParticleDataGroup:2022pth}.  In addition to the experimental search, theoretical studies have been conducted by numerous research groups to examine the spectroscopic characteristics of $B_c$ mesons.  For example, QCD sum rules~\cite{Dominguez:1993rg, Gershtein:1994dxw, Bagan:1994dy, Chen:2013eha, Wang:2012kw, Wang:2024fwc}, lattice QCD~\cite{Davies:1996gi, deDivitiis:2003iy, Allison:2004hy}, the heavy quark effective theory~\cite{Zeng:1994vj}, Blankenbecler-Sugar equation~\cite{Lahde:2002wj},  the quark  models~\cite{Akbar:2019kbi, Li:2019tbn, Asghar:2019qjl, Akbar:2018hiw, Monteiro:2016rzi, Li:2022bre, Li:2023wgq, Gao:2024yvz, Godfrey:1985xj, Chang:2019wpt, Hao:2024nqb}, and the Dyson-Schwinger equation~\cite{Chang:2019eob, Chen:2020ecu}. The predictions from these approaches can be validated once further experimental data on these mesons are obtained. Despite the significant advancements in heavy quark physics over recent years, the bottom-charm spectroscopy field remains relatively understudied, necessitating further research. In light of the ongoing advancements in experimental techniques, it is anticipated that the ATLAS, CMS, LHCb, and other relevant facilities will observe an increased number of $B_c$ mesons in the future.

A significant number of heavy hadrons remain to be discovered. Furthermore, the knowledge base concerning the properties of these hadrons is incomplete. The study of hadronic mass spectra provides crucial information for the experimental search for these particles, as well as important input parameters for the study of their properties.  Consequently, the study of the diverse properties of hadrons represents a crucial undertaking. 
This paper presents a calculation of the magnetic moment of the ground-state axial-vector and vector $B_c$ mesons within the framework of the QCD light-cone sum rules (LCSR). Magnetic moments represent a basic property intrinsic to every hadron. They play a significant role in the comprehension of the hadronic structure. In theoretical frameworks, the magnetic moment is linked to the magnetic form factor associated with the particular hadron under examination. 
A comprehensive understanding of the electromagnetic characteristics of $B_c$ mesons is essential for advancing our knowledge of confinement and heavy flavor effects.

The manuscript is arranged as follows. The basic procedure of LCSR for the magnetic moment of the $B_c$ mesons with $ \rm{J^{P}}= 1^{+}$ and $ \rm{J^{P}}= 1^{-}$ are discussed in detail in Section~\ref{formalism}.  In section~\ref{numerical} we present the numerical results and discussions.  In the last section, a brief conclusion is presented.

 \begin{widetext}
 
\section{Theoretical Frame}\label{formalism}

This section outlines the derivation of LCSR for ground-state axial-vector and vector $B_c$ mesons. To achieve this, it is required to consider the following correlation function within the presence of an external electromagnetic field ($F$), 
\begin{equation}
 \label{edmn01}
\Pi _{\alpha \beta }(p,q)=i\int d^{4}x\,e^{ip\cdot x}\langle 0|\mathcal{T} {J}_{\alpha}(x)
J_{\beta }^{\dagger }(0)\}|0\rangle_{F}, 
\end{equation}
where $J_{\alpha}(x)= J_{\alpha}^{A}(x)$, $J_{\alpha}^{V}(x)$; the axial-vector $J_{\alpha}^{A}(x)$() and vector ($J_{\alpha}^{V}(x)$) interpolating current  interpolate the axial-vector and vector $B_c$  mesons, respectively. As these parameters are of significance in further stages of the analysis, it is required that the explicit form of these currents be provided, as detailed below: 
\begin{align}
\label{curr1}
J_{\alpha }^{A}(x) &=  \bar c^a (x) \gamma_\alpha \gamma_5  b^{a} (x),\\
J_{\alpha }^{V}(x) &=  \bar c^a (x) \gamma_\alpha  b^{a} (x).\label{curr2}
\end{align}

In the context of the QCD light-cone sum rules method, the analysis process is conducted by the following steps:
\begin{itemize}
 \item In the initial stage, the pertinent correlation function is derived by hadronic observables, encompassing elements such as mass, decay constants, and other related quantities, which is designated as the "hadronic representation" or "hadronic side".
 
 \item The second step involves deriving the corresponding correlation function in terms of the quark-gluon degrees of freedom, a process known as the "QCD representation" or "QCD side".
 
 \item As a final step, the representations of the correlation function are matched with the support provided by the quark-hadron duality ansatz. The Borel transformations and continuum subtraction are employed to eliminate undesirable contributions from the calculations and to derive the sum rules for the physical parameter, which in this case is the magnetic moment,  to be calculated.
\end{itemize}

We may now proceed with our analysis in a manner consistent with the aforementioned steps. The initial step will be to derive the relevant physical quantity in terms of hadron parameters.  
In the hadronic representation, the correlation function is inserted with complete sets of hadronic states with the same
quantum numbers as $B_c$ mesons. Integration over x is performed to obtain the following representation of the correlation function of Eq. (\ref{edmn01}):
\begin{align}
\label{edmn04}
\Pi_{\alpha\beta}^{Had} (p,q) &= {\frac{\langle 0 \mid J_\alpha (x) \mid
B_c(p,\varepsilon^i) \rangle}{p^2 - m_{B_c}^2}} \langle B_c (p, \varepsilon^i) \mid B_c (p+q, \varepsilon^f) \rangle_F 
\frac{\langle B_c (p+q, \varepsilon^f) \mid J_{\beta }^{\dagger } (0) \mid 0 \rangle}{(p+q)^2 - m_{B_c}^2} \nonumber\\
&+ \mbox{higher states and continuum}.
\end{align}
  
  The meson vacuum matrix elements in the above correlation function can be expressed as follows: 
  \begin{align}
\label{edmn05}
\langle 0 \mid J_\alpha (x) \mid B_c (p, \varepsilon^i) \rangle &=  m_{B_c} f_{B_c} \varepsilon_\alpha^i\,,\\
\langle B_c (p+q, \varepsilon^{f}) \mid J_{\beta }^{\dagger } (0) \mid 0 \rangle &= m_{B_c} f_{B_c} \varepsilon_\beta^{* f}\,,
\end{align}
where $f_{B_c}$ and $ \varepsilon_\alpha^i (\varepsilon_\beta^{*f}) $   are the decay constant and the polarization vector of $B_c$ mesons, respectively. The remaining transition matrix element in Eq. (\ref{edmn04}) can be represented regarding the form factors for the spin-1 hadrons~\cite{Brodsky:1992px},
\begin{align}
\langle B_c(p,\varepsilon^i) \mid  B_c (p+q,\varepsilon^{f})\rangle_F &= - \varepsilon^\gamma (\varepsilon^{i})^\mu (\varepsilon^{f})^\nu
\Big[ G_1(Q^2)~ (2p+q)_\gamma ~g_{\mu\nu}  
+ G_2(Q^2)~ ( g_{\gamma\nu}~ q_\mu -  g_{\gamma\mu}~ q_\nu)
\nonumber\\ 
&
- \frac{1}{2 m_{B_c}^2} G_3(Q^2)~ (2p+q)_\gamma
q_\mu q_\nu  \Big],\label{edmn06}
\end{align}
where $\varepsilon^\gamma$ is the polarization vector of the photon, and  $G_i(Q^2)$'s are Lorentz invariant form factors of the corresponding transition. These form factors are a function of the $Q^2=-q^2$.   

In consideration of the aforementioned equations, the hadronic representation of the correlation function of the $B_c$ mesons is given by the following expression:
%
%
\begin{align}
\label{edmn09}
 \Pi_{\alpha\beta}^{Had}(p,q) &=  \frac{\varepsilon^\gamma \, m_{B_c}^2 f_{B_c}^2}{ [m_{B_c}^2 - (p+q)^2][m_{B_c}^2 - p^2]}
 \Bigg\{G_1(Q^2)(2p+q)_\gamma\Bigg[g_{\alpha\beta}-\frac{p_\alpha p_\beta}{m_{B_c}^2}
 -\frac{(p+q)_\alpha (p+q)_\beta}{m_{B_c}^2}+\frac{(p+q)_\alpha p_\beta}{2m_{B_c}^4}\nonumber\\
 & \times (Q^2+2m_{B_c}^2)
 \Bigg]
 + G_2 (Q^2) \Bigg[q_\alpha g_{\gamma\beta}  
 - q_\beta g_{\gamma\alpha}-
\frac{p_\beta}{m_{B_c}^2}  \big(q_\alpha p_\gamma - \frac{1}{2}
Q^2 g_{\alpha\gamma}\big) 
+
\frac{(p+q)_\alpha}{m_{B_c}^2}  \big(q_\beta (p+q)_\gamma+ \frac{1}{2}
Q^2 g_{\beta\gamma}\big) 
\nonumber\\
&-  
\frac{(p+q)_\alpha p_\beta p_\gamma}{m_{B_c}^4} \, Q^2
\Bigg]
-\frac{G_3(Q^2)}{m_{B_c}^2}(2p+q)_\gamma \Bigg[
q_\alpha q_\beta -\frac{p_\alpha q_\beta}{2 m_{B_c}^2} Q^2 
+\frac{(p+q)_\alpha q_\beta}{2 m_{B_c}^2} Q^2
-\frac{(p+q)_\alpha q_\beta}{4 m_{B_c}^4} Q^4\Bigg]
\Bigg\}\,.
\end{align}


In the static limit, defined by $Q^2 = 0$, the magnetic moment is illustrated by the following formula:
\begin{align}
\label{edmn08}
 \mu  &= \frac{e}{2 m_{B_c}}\,F_M(0), 
\end{align}
where $F_M(0) = G_2(0)$. 

The required equation for the magnetic moment has been derived, thereby paving the way for the construction of the hadronic representation of the analysis. The second step of the aforementioned process may now be commenced, namely the determination of the correlation function in terms of the parameters associated with quark-gluon degrees of freedom.

 
In the QCD representation of the correlation function, the interpolating currents of the $B_c$ mesons given in Eqs.~(\ref{curr1})-(\ref{curr2}) are inserted into the correlation function in Eq. (\ref{edmn01}). Following this step, Wick's theorem is employed to perform all pertinent contractions and derive the corresponding expressions, which are presented below as,
%
\begin{align}
 \label{eq:QCDSide}
\Pi _{\alpha \beta }^{\mathrm{QCD},\,A}(p,q)&=i \int d^{4}xe^{ip\cdot x} 
\langle 0 \mid \bigg\{ 
 \mathrm{Tr}
\Big[ \gamma _{\alpha } \gamma_5{S}_{b}^{aa^{\prime }}(x)\gamma _{\beta} \gamma_5 S_{c}^{a^{\prime }a}(-x)\Big]    
  \bigg\} \mid 0 \rangle_{F} , \\
\Pi _{\alpha \beta }^{\mathrm{QCD},\,V}(p,q)&=i \int d^{4}xe^{ip\cdot x} 
\langle 0 \mid \bigg\{ 
 \mathrm{Tr}
\Big[ \gamma _{\alpha }  {S}_{b}^{aa^{\prime }}(x)\gamma _{\beta}  S_{c}^{a^{\prime }a}(-x)\Big]    
  \bigg\} \mid 0 \rangle_{F},
  \label{eq:QCDSide2}
\end{align}%
where the heavy  quark propagator, $S_{Q}(x)$, is described as~\cite{Belyaev:1985wza}:
\begin{align}
S_{Q}(x)&=S_Q^{free}(x)
-\frac{m_{Q}\,g_{s}\, G^{\alpha \beta}(x)}{32\pi ^{2}} \bigg[ (\sigma _{\alpha \beta }{\xslash}
+{\xslash}\sigma _{\alpha \beta }) 
    \frac{K_{1}\big( m_{Q}\sqrt{-x^{2}}\big) }{\sqrt{-x^{2}}}
 +2\sigma_{\alpha \beta }K_{0}\big( m_{Q}\sqrt{-x^{2}}\big)\bigg],
 \label{edmn14}
\end{align}%
with  
\begin{align}
 S_Q^{free}(x)&=\frac{m_{Q}^{2}}{4 \pi^{2}} \bigg[ \frac{K_{1}\big(m_{Q}\sqrt{-x^{2}}\big) }{\sqrt{-x^{2}}}
+i\frac{{\xslash}~K_{2}\big( m_{Q}\sqrt{-x^{2}}\big)}
{(\sqrt{-x^{2}})^{2}}\bigg],
\end{align}
where $G^{\alpha\beta}$ is the gluon field-strength tensor,  and $K_i$'s are the Bessel functions.  Here, we use the following integral representation  of the Bessel function,     
\begin{equation}\label{b2}
K_n(m_Q\sqrt{-x^2})=\frac{\Gamma(n+ 1/2)~2^n}{m_Q^n \,\sqrt{\pi}}\int_0^\infty dt~\cos(m_Qt)\frac{(\sqrt{-x^2})^n}{(t^2-x^2)^{n+1/2}}.
\end{equation}

In the calculation of the magnetic moment of light quark-containing hadrons using the LCSR method, the aforementioned correlation functions encompass distinct contributions. These include the emission of a photon both perturbatively and non-perturbatively. However, in the context of $B_c$ mesons, which lack a valence light quark, the emission of a photon non-perturbatively is not a possibility. From a technical standpoint, the non-perturbative contributions are proportional to the product of distribution amplitudes and quark condensates. The heavy quark condensates are proportional to $1/m_Q$~\cite{Antonov:2012ud}. As a consequence of the large mass of the heavy quarks, such condensates for the heavy quarks will be largely suppressed. Consequently, non-perturbative contributions containing heavy quarks have not been included in our computations. Only the perturbative photon emission from the heavy quarks is taken into account.  

In order to include perturbative contributions into the analysis, it is necessary to implement the following replacement under the methodology, which is characterized as follows:
\begin{align}
\label{free}
S_Q^{free}(x) \longrightarrow \int d^4z\, S_Q^{free} (x-z)\,\rlap/{\!A}(z)\, S_Q^{free} (z)\,,
\end{align}
   where we use $ A_\mu(z)=-\frac{1}{2}\, F_{\mu\nu}(z)\, z^\nu $. Here  the electromagnetic field strength tensor is written as $ F_{\mu\nu}(z)=-i(\varepsilon_\mu q_\nu-\varepsilon_\nu q_\mu)\,e^{iq.z} $. 
   
Once the aforementioned modifications have been performed, namely when perturbative contributions are taken into account in the analysis, the QCD representation of the correlation function is obtained. 
After some mathematical manipulations for the $S_Q^{free}(x)$, its final form becomes:
\begin{eqnarray}\label{sfreepert}
&& S_Q^{free}(x)=-i\frac{e_Q m_Q}{32 \pi^2}
\bigg(\varepsilon_\alpha q_\beta-\varepsilon_\beta q_\alpha\bigg)
\bigg[2\sigma_{\alpha\beta}K_{0}\Big( m_{Q}\sqrt{-x^{2}}\Big)
 +\frac{K_{1}\Big( m_{Q}\sqrt{-x^{2}}\Big) }{\sqrt{-x^{2}}}
 \Big(\xslash\sigma_{\alpha \beta}+\sigma_{\alpha\beta}\xslash\Big)\bigg].
\end{eqnarray}

This amounts to taking $\bar T_4^{\gamma} (\underline{\alpha}) = 0$ and $S_{\gamma} (\underline {\alpha}) = \delta(\alpha_{\bar q})\delta(\alpha_{q})$ as the light-cone distribution amplitude in the three particle distribution amplitudes (see Ref. \cite{Li:2020rcg}). 
A comprehensive description of the procedures employed to encompass these contributions within the calculations can be found in Refs.~\cite{Ozdem:2022vip, Ozdem:2022eds}. 

%


After evaluating the correlation function for both the hadronic and the QCD representations, we proceed to equate the coefficient of the $(\varepsilon. p) (p_\alpha q_\beta-p_\beta q_\alpha)$ structure in both representations. Following this step, we perform a double Borel transformation for $p^2 $ and $(p + q)^2$. To account for the continuum, the quark-hadron duality ansatz is employed. This ultimately yields the LCSR for the magnetic moments of the ground-state $B_c$ mesons, which can be expressed as follows:
\begin{align}
\label{avector}
 \mu_{B_c}^{A}\,  &=  \, \frac{e^{\frac{m_{B_c}}{\rm{M^2}}}}{f^{2}_{B_c} m^{4}_{B_c}} \, \rho_1(\rm{M^2}, \rm{s_0}) , \\
  \mu_{B_c}^{V}  &=  \frac{e^{\frac{m_{B_c}}{\rm{M^2}}}}{f^{2}_{B_c} m^{4}_{B_c}} \,  \rho_2(\rm{M^2}, \rm{s_0}), \label{vector} 
 \end{align}
with 
\begin{align}
\label{axial1}
 \rho_1(\rm{M^2}, \rm{s_0}) &= \frac{3 (e_b-e_c)}{32 \pi} \Bigg[  \,
\int_{\mathcal M}^{\rm{s_0}} ds \, \int_0^1 dt~ e^{-s/\rm{M^2}}~ 
 (s-\mathcal M)^3 \, t^2
 +  
 \int_{\mathcal M}^{\rm{s_0}} ds \, \int_0^1 dt~ e^{-s/\rm{M^2}}~ 
 (s-\mathcal M)^3 \, t^3\nonumber\\
 &
 +
 \int_{\mathcal M}^{\rm{s_0}} ds \, \int_0^1 dt~ e^{-s/\rm{M^2}}~ 
 (s-\mathcal M)^3 \, t^4
 +
 \int_{\mathcal M}^{\rm{s_0}} ds \, \int_0^1 dt~ e^{-s/\rm{M^2}}~ 
 (s-\mathcal M)^3 \, t^5 \Bigg],
\end{align}
\begin{align}
\label{vector1}
 \rho_2(\rm{M^2}, \rm{s_0})&= -\frac{3 (e_b-e_c)}{32 \pi} \Bigg[
\int_{\mathcal M}^{\rm{s_0}} ds \, \int_0^1 dt~ e^{-s/\rm{M^2}}~ 
 (s-\mathcal M)^3 \, t^2
 +  
 \int_{\mathcal M}^{\rm{s_0}} ds \, \int_0^1 dt~ e^{-s/\rm{M^2}}~ 
 (s-\mathcal M)^3 \, t^3\nonumber\\
 &
 +
 \int_{\mathcal M}^{\rm{s_0}} ds \, \int_0^1 dt~ e^{-s/\rm{M^2}}~ 
 (s-\mathcal M)^3 \, t^4
 +
 \int_{\mathcal M}^{\rm{s_0}} ds \, \int_0^1 dt~ e^{-s/\rm{M^2}}~ 
 (s-\mathcal M)^3 \, t^5 \Bigg],
\end{align}

\noindent   where  $\mathcal M = (m_c+m_b)^2$.

The analytical formulae for the magnetic moments of $B_c$ mesons are provided here. The following section will detail the numerical computations of these characteristics.\end{widetext}

\section{Numerical evaluations}\label{numerical}

In order to determine magnetic moments, numerical calculations of the LCSR are required, which necessitate the input of some variables; these values can be found in Table~\ref{inputparameter}.    
  \begin{table}[htp]
	\addtolength{\tabcolsep}{10pt}
	\caption{ Input parameters used in calculations~\cite{Workman:2022ynf, Wang:2012kw}.}
	\label{inputparameter}
\begin{tabular}{l|c|ccccc}
               \hline\hline
Parameter & Value&Unit \\
                                        \hline\hline
$m_c$& $ 1.27 \pm 0.02$& GeV                 \\
$m_b$& $ 4.18^{+0.03}_{-0.02}$& GeV                    \\
$m_{B_c}(1^+)$& $  6.730 \pm 0.061 $& GeV                       \\
$m_{B_c} (1^-)$& $  6.337 \pm 0.052 $& GeV                       \\                    
$f_{B_c}(1^+)$& $  0.373 \pm 0.025  $& GeV  \\
$f_{B_c}(1^-)$&$  0.384 \pm 0.032    $& GeV  \\                                    \hline\hline
 \end{tabular}
\end{table}

In order to conduct a more detailed analysis, two additional parameters are required in conjunction with the aforementioned input variables. These are the Borel mass parameters, denoted by $\rm{M^2}$, and the continuum threshold parameter, denoted by $\rm{s_0}$. 
In order to achieve accurate and reliable results from the LCSR, it is essential to delineate the region where the dependence of the magnetic moments on these parameters is relatively weak. This region is referred to as the "working window" or "working region".  
 The functionality of these auxiliary parameters is constrained by the established methodology, which adheres to two core principles: pole dominance (PC) and convergence of OPE (CVG).   
%
As our analysis is solely based on perturbative contributions, a CVG analysis is not within the scope of this study. Instead, it is sufficient to determine the PC limitation, for which the relevant formula is provided below:
\begin{align}
 \mbox{PC} &=\frac{\rho_i (\rm{M^2},\rm{s_0})}{\rho_i (\rm{M^2},\infty)},
 \end{align}
 where  $\rho_i (\rm{M^2},\rm{s_0})$ are given in Eqs.~(\ref{axial1}) and (\ref{vector1}).
 
\begin{figure}[htp]
\centering
  \subfloat[]{\includegraphics[width=0.4\textwidth]{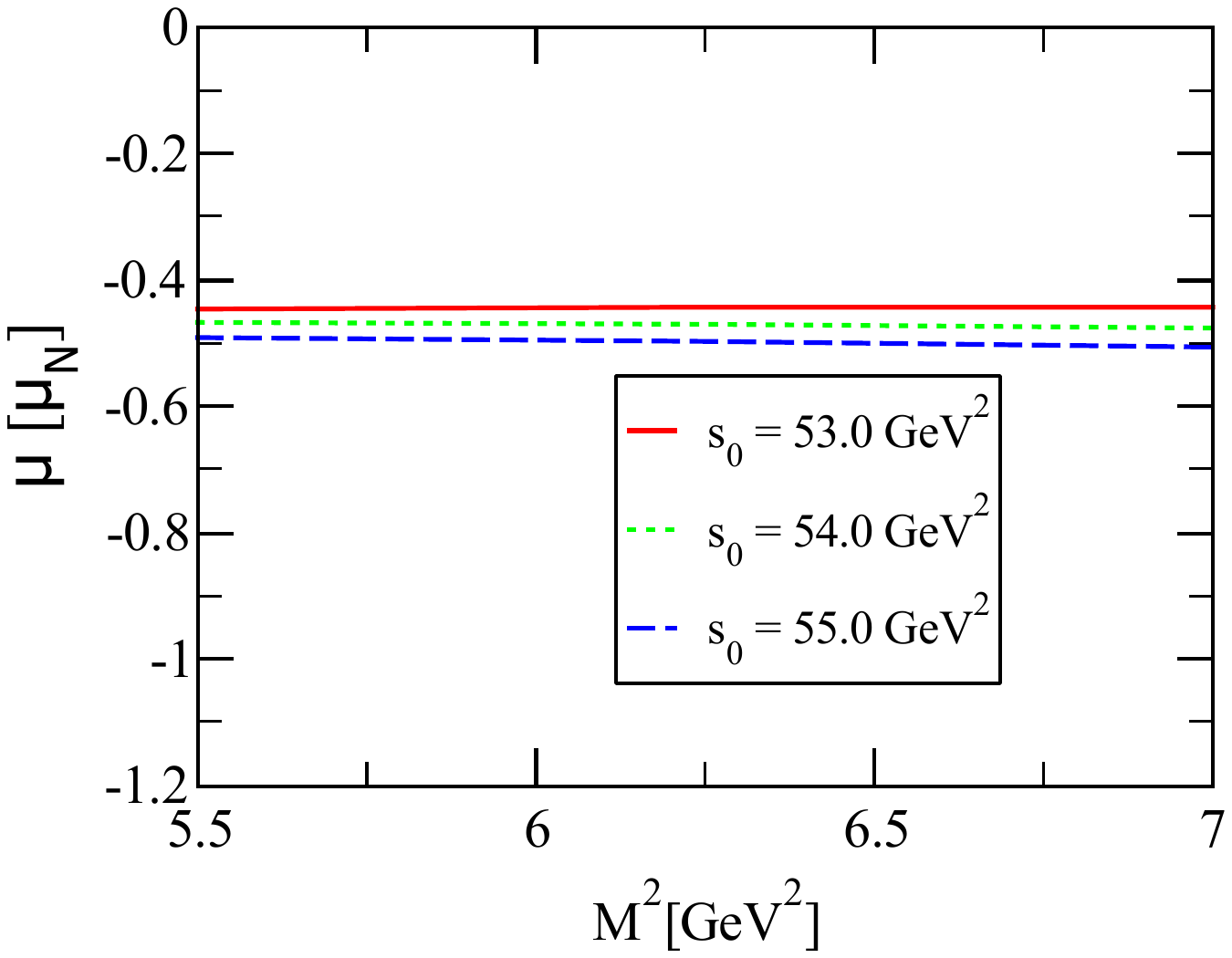}} ~~~~
  \subfloat[]{\includegraphics[width=0.4\textwidth]{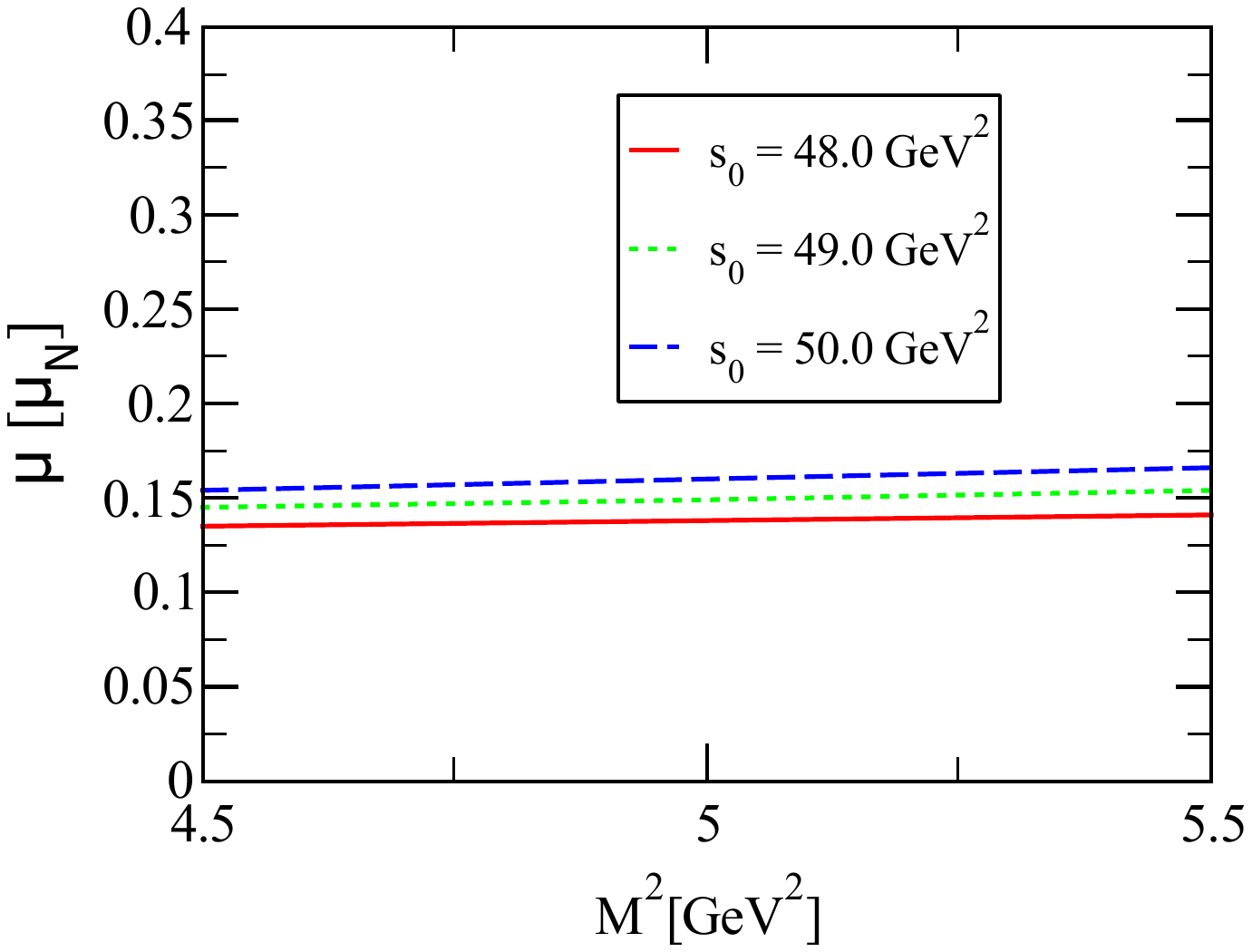}} \\
  \caption{ Variation of magnetic moments $B_c$ mesons as a function of the $\rm{M^2}$ at different values of $\rm{s_0}$: (a) and (b) are for axial-vector and vector states, respectively.}
 \label{figMsq}
  \end{figure}

 By the aforementioned stipulation, the working windows of the auxiliary parameters, as delineated in Table \ref{table}, are derived.   
 As demonstrated by the outcomes, the methodology aligns with the constraints outlined in the specifications.  
 In order to enhance the precision of our predictions and ensure their completeness, Fig.~\ref{figMsq} illustrates the variations in the derived magnetic moments of $B_c$ mesons concerning auxiliary parameters.  As illustrated in the figure, the magnetic moments of $B_c$ mesons demonstrate a relatively modest variation  with these auxiliary variables. The requisite criteria inherent to the LCSR have now been satisfied, and it seems reasonable to anticipate reliable predictions. 
 %

\begin{figure}[htp]
\centering
  \subfloat[]{\includegraphics[width=0.43\textwidth]{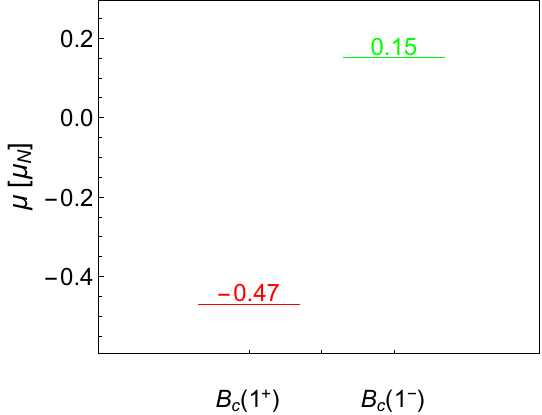}} ~~~~
  \subfloat[]{\includegraphics[width=0.43\textwidth]{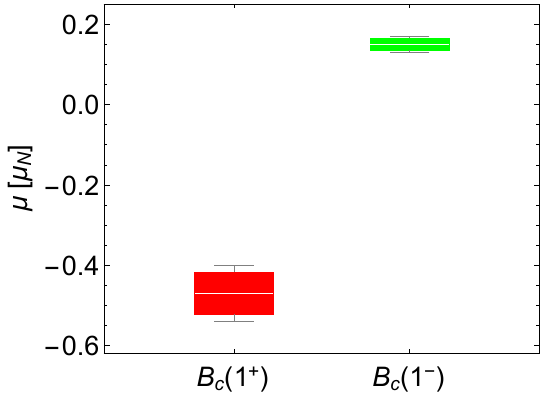}} \\
  \caption{ The magnetic moments of $B_c$ mesons: (a) for central values, and (b) for combined with errors, respectively.}
 \label{figMsq1}
  \end{figure}

   Now that all the requisite parameters for the magnetic moments of the axial-vector and vector $B_c$ mesons have been identified, the necessary numerical analysis can be undertaken. To assist the reader in visualizing and comprehending the results, we have provided them in Fig. \ref{figMsq1}. The magnetic moment results obtained from the numerical analysis are presented in Table \ref{table}, in which all the uncertainties in the input parameters have been taken into account.
  %
  \begin{table}[htb!]
	\addtolength{\tabcolsep}{10pt}
	\caption{The predicted magnetic moments of the vector and axial-vector $B_c$ mesons.}
	\label{table}
	\begin{center}
\begin{tabular}{lccccccc}
	   \hline\hline
	   \\
	   States &  $ \rm{J^{P}}$ &$\mu$\,\,[$\rm{\mu_N}$]	& $\rm{M^2}\,\,[\rm{GeV}^2]$& $\rm{s_0}\,\,[\rm{GeV}^2]$& PC\,\,[$\%$] 	   \\
	   \\
	   \hline\hline
	  \\
 $B_c$&$1^{+}$&   $ -0.47 \pm 0.07$      & [5.5, 7.0]& [53.0, 55.0]& [68.05, 51.10]\\
	   \\
   
$B_c$&$1^{-}$& $ ~~0.15 \pm 0.02$          & [4.5, 5.5]& [48.0, 50.0]& [65.21, 51.18]\\
	    \\
	   \hline\hline
\end{tabular}
\end{center}
\end{table}
 
\begin{figure}[htp]
\centering
  \includegraphics[width=0.44\textwidth]{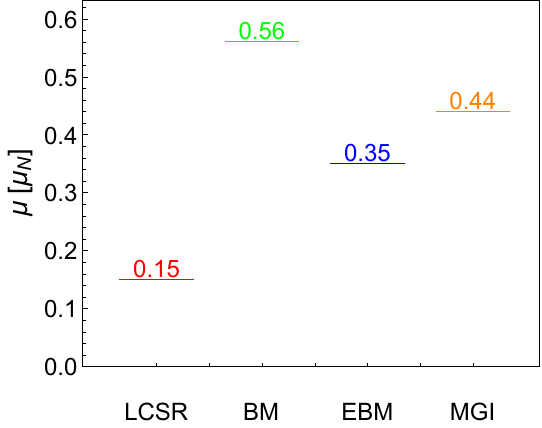} 
  \caption{ The comparison of the magnetic moment of the vector $B_c$ meson with BM~\cite{Bose:1980vy}, EBM~\cite{Simonis:2016pnh}, 
  and MGI~\cite{Li:2023wgq}.}
 \label{figMsq2}
  \end{figure}

At the end, we would like to present a comparison of our result on the vector $B_c$ meson magnetic
moment, with the ones existing in the literature. A review of the literature reveals that there are no results for the axial-vector $B_c$ meson; thus, a comparison cannot be made. 
 The magnetic moments of the ground-state vector $B_c$  meson were obtained utilizing the Bag model (BM)~\cite{Bose:1980vy}, extended Bag model (EBM) \cite{Simonis:2016pnh},   
 and modified Godfrey-Isgur model (MGI)~\cite{Li:2023wgq}.  The obtained results are given as $\mu_{B_c} = 0.56\, \mu_N $,  $\mu_{B_c} = 0.35\, \mu_N $, and $\mu_{B_c} = 0.44\, \mu_N $ for the Bag model, extended Bag model, and modified Godfrey-Isgur model, respectively.   To facilitate a more comprehensive understanding of the comparison, the results have also been presented in Fig. \ref{figMsq2}.
 %
 As can be seen from these predictions, the magnetic moment of the vector $B_c$ meson has been found to vary considerably between different theoretical models, which provides a means of distinguishing between them.  The discrepancy between the results of different approaches may be attributed to the choice of wave functions employed in each. However, the source of this discrepancy remains unclear. Further theoretical and experimental studies are required to elucidate the inconsistencies and gain a deeper understanding of the current situation. However, direct measurements of the magnetic moment of the vector $B_c$ meson are not yet feasible. Consequently, any indirect projections of the magnetic moment of the vector $B_c$ meson would be valuable. 
 %
%
 %
%

It is important to briefly discuss the methodology for measuring the magnetic moment of the $B_c$ mesons in an experiment before concluding this analysis. Measuring the magnetic moments of the $B_c$ mesons employing the spin precession or vector-meson-electron scattering experiments is difficult as a consequence of their short lifetimes.  The magnetic moment of short-lived hadrons can only be determined indirectly through a three-step process. In the initial stage of the process, the corresponding hadron is produced. Subsequently, the hadron emits a photon of low energy, which acts as an external magnetic field. In the final analysis, the particle undergoes a decay process.  Alternative methodologies have been proposed for the indirect estimation of the magnetic moments of vector/axial-vector mesons through the application of the aforementioned technique.  The initial approach is founded upon the hypothesis of soft photon emission by hadrons, as proposed in Ref.~\cite{Zakharov:1968fb}. This theory outlines a methodology for the calculation of electromagnetic multipole moments.  
The fundamental premise of this technique is that the photon carries data regarding the magnetic dipole and higher multipole moments of the emitting particle. 
The matrix element for the corresponding radiative process is written as
\begin{align}
 M \sim A \,w_\gamma^{-1} + B\,w_\gamma^0 + \cdots,
\end{align}
 where $w_\gamma$ is the energy of the photon.  The electric charge contributes to the amplitude at order $w_\gamma^ {-1}$, while the contribution from the magnetic moment is proportional to $w_\gamma^0$. The higher multipole moments are denoted by the $\cdots$.   Accordingly, the magnetic moment of the hadrons under discussion can be determined by measuring the cross-section or decay width of the radiative process, while the minor effects of the linear/higher-order terms in $w_\gamma$ can be ignored.  The magnetic moment of the $ \Delta(1232)$ baryon was extracted utilizing the application of this technique~~\cite{Pascalutsa:2004je, Pascalutsa:2005vq, Pascalutsa:2007wb,   Kotulla:2002cg, Drechsel:2001qu, Machavariani:1999fr, Drechsel:2000um, Chiang:2004pw, Machavariani:2005vn}.   
 The second one is the possibility of measuring the electromagnetic characteristics of the vector mesons in the radiative production and decays of such mesons, and it was proposed that the energy and the combined angular distributions of the radiated photons is an effective method to measure the electromagnetic characteristics of the vector mesons~\cite{LopezCastro:1997dg}. This technique has been employed to ascertain the magnetic moment of the $\rho$ meson, with preliminary data from the BaBar Collaboration about the $e^+ e^- \rightarrow \pi^+ \pi^- 2 \pi^0$ process, in the center of mass energy range from $0.9$ to $2.2$ GeV and obtained $\mu_\rho = 2.1 \pm 0.5~e/{2 m_\rho}$ units for the magnetic moment of the $\rho$ meson~\cite{GarciaGudino:2013alv}.  Hence, while it is at present not feasible to directly measure the electromagnetic characteristics of short-life hadrons through experimentation, they can be extracted through the indirect analysis of the data collected from corresponding radiative processes.

\section{Conclusion}\label{conc}

The magnetic moments of the ground-state $B_c$ mesons provide significant insights into their inner structure and geometric shape. Furthermore, a comprehensive understanding of the electromagnetic characteristics of ground-state $B_c$ mesons is essential for advancing our knowledge of confinement and heavy flavor effects. In light of this, we proceed to extract the magnetic moments of the ground-state vector and axial-vector $B_c$ mesons through the medium of the QCD light-cone sum rules. The magnetic moments of the axial-vector and vector $B_c$ mesons are found to be $\mu_{B_c}= -0.47 \pm 0.07~\mu_N$, and $\mu_{B_c}= 0.15 \pm 0.02~\mu_N$, respectively. A comparison of our results for the vector $B_c$ meson with other theoretical predictions has revealed discrepancies between the various predictions, which could prove useful as a complementary tool for interpreting the vector $B_c$ meson. The current experimental data set is limited to a small number of observed states of beauty-charm mesons. However, theoretical studies can play a valuable role in elucidating their nature and guiding future experimental investigations.
   
 \section{Acknowledgments} 
The author would like to acknowledge A. Özpineci for his invaluable contributions to the comments, discussions, and suggestions presented in this work.
 
\bibliography{Bcmesons.bib}

\begin{thebibliography}{10}
\expandafter\ifx\csname url\endcsname\relax
  \def\url#1{\texttt{#1}}\fi
\expandafter\ifx\csname urlprefix\endcsname\relax\def\urlprefix{URL }\fi
\expandafter\ifx\csname href\endcsname\relax
  \def\href#1#2{#2} \def\path#1{#1}\fi

\bibitem{CDF:1998ihx}
F.~Abe, et~al., {Observation of the $B_c$ meson in $p\bar{p}$ collisions at
  $\sqrt{s} = 1.8$ TeV}, Phys. Rev. Lett. 81 (1998) 2432--2437.
\newblock \href {http://arxiv.org/abs/hep-ex/9805034}
  {\path{arXiv:hep-ex/9805034}}, \href
  {https://doi.org/10.1103/PhysRevLett.81.2432}
  {\path{doi:10.1103/PhysRevLett.81.2432}}.

\bibitem{CDF:2005yjh}
A.~Abulencia, et~al., {Evidence for the exclusive decay $B_c^\pm \to J/\psi
  \pi^\pm$ and measurement of the mass of the $B_c$ meson}, Phys. Rev. Lett. 96
  (2006) 082002.
\newblock \href {http://arxiv.org/abs/hep-ex/0505076}
  {\path{arXiv:hep-ex/0505076}}, \href
  {https://doi.org/10.1103/PhysRevLett.96.082002}
  {\path{doi:10.1103/PhysRevLett.96.082002}}.

\bibitem{CDF:2007umr}
T.~Aaltonen, et~al., {Observation of the Decay $B^+$ -($c$) $\to J/\psi
  \pi^\pm$ and Measurement of the $B^+$ -($c$) Mass}, Phys. Rev. Lett. 100
  (2008) 182002.
\newblock \href {http://arxiv.org/abs/0712.1506} {\path{arXiv:0712.1506}},
  \href {https://doi.org/10.1103/PhysRevLett.100.182002}
  {\path{doi:10.1103/PhysRevLett.100.182002}}.

\bibitem{D0:2008bqs}
V.~M. Abazov, et~al., {Observation of the $B_c$ Meson in the Exclusive Decay
  $B_c \to J/\psi \pi$}, Phys. Rev. Lett. 101 (2008) 012001.
\newblock \href {http://arxiv.org/abs/0802.4258} {\path{arXiv:0802.4258}},
  \href {https://doi.org/10.1103/PhysRevLett.101.012001}
  {\path{doi:10.1103/PhysRevLett.101.012001}}.

\bibitem{LHCb:2012ag}
R.~Aaij, et~al., {First observation of the decay $B_c^+ \to J/\psi
  \pi^+\pi^-\pi^+$}, Phys. Rev. Lett. 108 (2012) 251802.
\newblock \href {http://arxiv.org/abs/1204.0079} {\path{arXiv:1204.0079}},
  \href {https://doi.org/10.1103/PhysRevLett.108.251802}
  {\path{doi:10.1103/PhysRevLett.108.251802}}.

\bibitem{LHCb:2013vrl}
R.~Aaij, et~al., {Observation of the decay $B_c^+ \to \psi(2S)\pi^+$}, Phys.
  Rev. D 87 (2013) 071103.
\newblock \href {http://arxiv.org/abs/1303.1737} {\path{arXiv:1303.1737}},
  \href {https://doi.org/10.1103/PhysRevD.87.071103}
  {\path{doi:10.1103/PhysRevD.87.071103}}.

\bibitem{LHCb:2013kwl}
R.~Aaij, et~al., {Observation of $B^+_c \rightarrow J/\psi D_s^+$ and $B^+_c
  \rightarrow J/\psi D_s^{*+}$ decays}, Phys. Rev. D 87~(11) (2013) 112012,
  [Addendum: Phys.Rev.D 89, 019901 (2014)].
\newblock \href {http://arxiv.org/abs/1304.4530} {\path{arXiv:1304.4530}},
  \href {https://doi.org/10.1103/PhysRevD.87.112012}
  {\path{doi:10.1103/PhysRevD.87.112012}}.

\bibitem{LHCb:2013hwj}
R.~Aaij, et~al., {First observation of the decay $B_{c}^{+}\to J/\psi K^+$},
  JHEP 09 (2013) 075.
\newblock \href {http://arxiv.org/abs/1306.6723} {\path{arXiv:1306.6723}},
  \href {https://doi.org/10.1007/JHEP09(2013)075}
  {\path{doi:10.1007/JHEP09(2013)075}}.

\bibitem{LHCb:2013rud}
R.~Aaij, et~al., {Observation of the decay $B_c \rightarrow J/\psi K^+ K^-
  \pi^+ $}, JHEP 11 (2013) 094.
\newblock \href {http://arxiv.org/abs/1309.0587} {\path{arXiv:1309.0587}},
  \href {https://doi.org/10.1007/JHEP11(2013)094}
  {\path{doi:10.1007/JHEP11(2013)094}}.

\bibitem{ATLAS:2014lga}
G.~Aad, et~al., {Observation of an Excited $B_c^\pm$ Meson State with the ATLAS
  Detector}, Phys. Rev. Lett. 113~(21) (2014) 212004.
\newblock \href {http://arxiv.org/abs/1407.1032} {\path{arXiv:1407.1032}},
  \href {https://doi.org/10.1103/PhysRevLett.113.212004}
  {\path{doi:10.1103/PhysRevLett.113.212004}}.

\bibitem{CMS:2019uhm}
A.~M. Sirunyan, et~al., {Observation of Two Excited B$^+_\mathrm{c}$ States and
  Measurement of the B$^+_\mathrm{c}$(2S) Mass in pp Collisions at $\sqrt{s} =$
  13 TeV}, Phys. Rev. Lett. 122~(13) (2019) 132001.
\newblock \href {http://arxiv.org/abs/1902.00571} {\path{arXiv:1902.00571}},
  \href {https://doi.org/10.1103/PhysRevLett.122.132001}
  {\path{doi:10.1103/PhysRevLett.122.132001}}.

\bibitem{LHCb:2019bem}
R.~Aaij, et~al., {Observation of an excited $B_c^+$ state}, Phys. Rev. Lett.
  122~(23) (2019) 232001.
\newblock \href {http://arxiv.org/abs/1904.00081} {\path{arXiv:1904.00081}},
  \href {https://doi.org/10.1103/PhysRevLett.122.232001}
  {\path{doi:10.1103/PhysRevLett.122.232001}}.

\bibitem{ParticleDataGroup:2022pth}
R.~L. Workman, et~al., {Review of Particle Physics}, PTEP 2022 (2022) 083C01.
\newblock \href {https://doi.org/10.1093/ptep/ptac097}
  {\path{doi:10.1093/ptep/ptac097}}.

\bibitem{Dominguez:1993rg}
C.~A. Dominguez, K.~Schilcher, Y.~L. Wu, {QCD determination of the leptonic
  decay constant of the B(c) meson}, Phys. Lett. B 298 (1993) 190--194.
\newblock \href {https://doi.org/10.1016/0370-2693(93)91729-7}
  {\path{doi:10.1016/0370-2693(93)91729-7}}.

\bibitem{Gershtein:1994dxw}
S.~S. Gershtein, V.~V. Kiselev, A.~K. Likhoded, A.~V. Tkabladze, {B(c)
  spectroscopy}, Phys. Rev. D 51 (1995) 3613--3627.
\newblock \href {http://arxiv.org/abs/hep-ph/9406339}
  {\path{arXiv:hep-ph/9406339}}, \href
  {https://doi.org/10.1103/PhysRevD.51.3613}
  {\path{doi:10.1103/PhysRevD.51.3613}}.

\bibitem{Bagan:1994dy}
E.~Bagan, H.~G. Dosch, P.~Gosdzinsky, S.~Narison, J.~M. Richard, {Hadrons with
  charm and beauty}, Z. Phys. C 64 (1994) 57--72.
\newblock \href {http://arxiv.org/abs/hep-ph/9403208}
  {\path{arXiv:hep-ph/9403208}}, \href {https://doi.org/10.1007/BF01557235}
  {\path{doi:10.1007/BF01557235}}.

\bibitem{Chen:2013eha}
W.~Chen, T.~G. Steele, S.-L. Zhu, {Masses of the bottom-charm hybrid $\bar bGc$
  states}, J. Phys. G 41 (2014) 025003.
\newblock \href {http://arxiv.org/abs/1306.3486} {\path{arXiv:1306.3486}},
  \href {https://doi.org/10.1088/0954-3899/41/2/025003}
  {\path{doi:10.1088/0954-3899/41/2/025003}}.

\bibitem{Wang:2012kw}
Z.-G. Wang, {Analysis of the vector and axialvector $B_c$ mesons with QCD sum
  rules}, Eur. Phys. J. A 49 (2013) 131.
\newblock \href {http://arxiv.org/abs/1203.6252} {\path{arXiv:1203.6252}},
  \href {https://doi.org/10.1140/epja/i2013-13131-7}
  {\path{doi:10.1140/epja/i2013-13131-7}}.

\bibitem{Wang:2024fwc}
Z.-G. Wang, {B $_{c}$ meson and its scalar cousin with QCD sum rules*}, Chin.
  Phys. C 48~(10) (2024) 103104.
\newblock \href {http://arxiv.org/abs/2401.12571} {\path{arXiv:2401.12571}},
  \href {https://doi.org/10.1088/1674-1137/ad5a71}
  {\path{doi:10.1088/1674-1137/ad5a71}}.

\bibitem{Davies:1996gi}
C.~T.~H. Davies, K.~Hornbostel, G.~P. Lepage, A.~J. Lidsey, J.~Shigemitsu,
  J.~H. Sloan, {B(c) spectroscopy from lattice QCD}, Phys. Lett. B 382 (1996)
  131--137.
\newblock \href {http://arxiv.org/abs/hep-lat/9602020}
  {\path{arXiv:hep-lat/9602020}}, \href
  {https://doi.org/10.1016/0370-2693(96)00650-8}
  {\path{doi:10.1016/0370-2693(96)00650-8}}.

\bibitem{deDivitiis:2003iy}
G.~M. de~Divitiis, M.~Guagnelli, R.~Petronzio, N.~Tantalo, F.~Palombi, {Heavy
  quark masses in the continuum limit of quenched lattice QCD}, Nucl. Phys. B
  675 (2003) 309--332.
\newblock \href {http://arxiv.org/abs/hep-lat/0305018}
  {\path{arXiv:hep-lat/0305018}}, \href
  {https://doi.org/10.1016/j.nuclphysb.2003.10.001}
  {\path{doi:10.1016/j.nuclphysb.2003.10.001}}.

\bibitem{Allison:2004hy}
I.~F. Allison, C.~T.~H. Davies, A.~Gray, A.~S. Kronfeld, P.~B. Mackenzie, J.~N.
  Simone, {A Precise determination of the $B_c$ mass from dynamical lattice
  QCD}, Nucl. Phys. B Proc. Suppl. 140 (2005) 440--442.
\newblock \href {http://arxiv.org/abs/hep-lat/0409090}
  {\path{arXiv:hep-lat/0409090}}, \href
  {https://doi.org/10.1016/j.nuclphysbps.2004.11.375}
  {\path{doi:10.1016/j.nuclphysbps.2004.11.375}}.

\bibitem{Zeng:1994vj}
J.~Zeng, J.~W. Van~Orden, W.~Roberts, {Heavy mesons in a relativistic model},
  Phys. Rev. D 52 (1995) 5229--5241.
\newblock \href {http://arxiv.org/abs/hep-ph/9412269}
  {\path{arXiv:hep-ph/9412269}}, \href
  {https://doi.org/10.1103/PhysRevD.52.5229}
  {\path{doi:10.1103/PhysRevD.52.5229}}.

\bibitem{Lahde:2002wj}
T.~A. Lahde, {Exchange current operators and electromagnetic dipole transitions
  in heavy quarkonia}, Nucl. Phys. A 714 (2003) 183--212.
\newblock \href {http://arxiv.org/abs/hep-ph/0208110}
  {\path{arXiv:hep-ph/0208110}}, \href
  {https://doi.org/10.1016/S0375-9474(02)01362-3}
  {\path{doi:10.1016/S0375-9474(02)01362-3}}.

\bibitem{Akbar:2019kbi}
N.~Akbar, {Properties of $B_c$ Mesons and Variational Constraints on Their
  Masses}, Phys. Atom. Nucl. 83~(4) (2020) 634--640.
\newblock \href {http://arxiv.org/abs/1911.02078} {\path{arXiv:1911.02078}},
  \href {https://doi.org/10.1134/S1063778820040031}
  {\path{doi:10.1134/S1063778820040031}}.

\bibitem{Li:2019tbn}
Q.~Li, M.-S. Liu, L.-S. Lu, Q.-F. L\"u, L.-C. Gui, X.-H. Zhong, {Excited
  bottom-charmed mesons in a nonrelativistic quark model}, Phys. Rev. D 99~(9)
  (2019) 096020.
\newblock \href {http://arxiv.org/abs/1903.11927} {\path{arXiv:1903.11927}},
  \href {https://doi.org/10.1103/PhysRevD.99.096020}
  {\path{doi:10.1103/PhysRevD.99.096020}}.

\bibitem{Asghar:2019qjl}
I.~Asghar, F.~Akram, B.~Masud, M.~A. Sultan, {Properties of excited
  charmed-bottom mesons}, Phys. Rev. D 100~(9) (2019) 096002.
\newblock \href {http://arxiv.org/abs/1910.02680} {\path{arXiv:1910.02680}},
  \href {https://doi.org/10.1103/PhysRevD.100.096002}
  {\path{doi:10.1103/PhysRevD.100.096002}}.

\bibitem{Akbar:2018hiw}
N.~Akbar, F.~Akram, B.~Masud, M.~Atif~Sultan, {Conventional and hybrid B$_{c}$
  mesons in an extended potential model}, Eur. Phys. J. A 55~(5) (2019) 82.
\newblock \href {http://arxiv.org/abs/1811.07552} {\path{arXiv:1811.07552}},
  \href {https://doi.org/10.1140/epja/i2019-12735-1}
  {\path{doi:10.1140/epja/i2019-12735-1}}.

\bibitem{Monteiro:2016rzi}
A.~P. Monteiro, M.~Bhat, K.~B. Vijaya~Kumar, {$c\bar{b}$ spectrum and decay
  properties with coupled channel effects}, Phys. Rev. D 95~(5) (2017) 054016.
\newblock \href {http://arxiv.org/abs/1608.05782} {\path{arXiv:1608.05782}},
  \href {https://doi.org/10.1103/PhysRevD.95.054016}
  {\path{doi:10.1103/PhysRevD.95.054016}}.

\bibitem{Li:2022bre}
T.-y. Li, L.~Tang, Z.-y. Fang, C.-h. Wang, C.-q. Pang, X.~Liu, {Higher states
  of the Bc meson family}, Phys. Rev. D 108~(3) (2023) 034019.
\newblock \href {http://arxiv.org/abs/2204.14258} {\path{arXiv:2204.14258}},
  \href {https://doi.org/10.1103/PhysRevD.108.034019}
  {\path{doi:10.1103/PhysRevD.108.034019}}.

\bibitem{Li:2023wgq}
X.-J. Li, Y.-S. Li, F.-L. Wang, X.~Liu, {Spectroscopic survey of higher-lying
  states of $B_c$ meson family}, Eur. Phys. J. C 83~(11) (2023) 1080.
\newblock \href {http://arxiv.org/abs/2308.07206} {\path{arXiv:2308.07206}},
  \href {https://doi.org/10.1140/epjc/s10052-023-12237-9}
  {\path{doi:10.1140/epjc/s10052-023-12237-9}}.

\bibitem{Gao:2024yvz}
Z.-b. Gao, Y.-y. Fan, H.~Chen, C.-q. Pang, {M1 radiative and spin-nonflip
  \ensuremath{\pi}\ensuremath{\pi} transitions of Bc states in the Cornell
  potential model}, Phys. Rev. D 110~(3) (2024) 034003.
\newblock \href {http://arxiv.org/abs/2402.10629} {\path{arXiv:2402.10629}},
  \href {https://doi.org/10.1103/PhysRevD.110.034003}
  {\path{doi:10.1103/PhysRevD.110.034003}}.

\bibitem{Godfrey:1985xj}
S.~Godfrey, N.~Isgur, {Mesons in a Relativized Quark Model with
  Chromodynamics}, Phys. Rev. D 32 (1985) 189--231.
\newblock \href {https://doi.org/10.1103/PhysRevD.32.189}
  {\path{doi:10.1103/PhysRevD.32.189}}.

\bibitem{Chang:2019wpt}
L.~Chang, M.~Chen, X.-q. Li, Y.-x. Liu, K.~Raya, {Can the Hyperfine Mass
  Splitting Formula in Heavy Quarkonia be Applied to the $B_c$ System?}, Few
  Body Syst. 62~(1) (2021) 4.
\newblock \href {http://arxiv.org/abs/1912.08339} {\path{arXiv:1912.08339}},
  \href {https://doi.org/10.1007/s00601-020-01586-w}
  {\path{doi:10.1007/s00601-020-01586-w}}.

\bibitem{Hao:2024nqb}
W.~Hao, R.~Zhu, {Beauty-charm Meson Family with Coupled Channel Effects and
  Their Strong Decays} (2 2024).
\newblock \href {http://arxiv.org/abs/2402.18898} {\path{arXiv:2402.18898}}.

\bibitem{Chang:2019eob}
L.~Chang, M.~Chen, Y.-x. Liu, {Excited $B_c$ states via the Dyson-Schwinger
  equation approach of QCD}, Phys. Rev. D 102~(7) (2020) 074010.
\newblock \href {http://arxiv.org/abs/1904.00399} {\path{arXiv:1904.00399}},
  \href {https://doi.org/10.1103/PhysRevD.102.074010}
  {\path{doi:10.1103/PhysRevD.102.074010}}.

\bibitem{Chen:2020ecu}
M.~Chen, L.~Chang, Y.-x. Liu, {$B_c$ meson spectrum via Dyson-Schwinger
  equation and Bethe-Salpeter equation approach}, Phys. Rev. D 101~(5) (2020)
  056002.
\newblock \href {http://arxiv.org/abs/2001.00161} {\path{arXiv:2001.00161}},
  \href {https://doi.org/10.1103/PhysRevD.101.056002}
  {\path{doi:10.1103/PhysRevD.101.056002}}.

\bibitem{Brodsky:1992px}
S.~J. Brodsky, J.~R. Hiller, {Universal properties of the electromagnetic
  interactions of spin one systems}, Phys. Rev. D 46 (1992) 2141--2149.
\newblock \href {https://doi.org/10.1103/PhysRevD.46.2141}
  {\path{doi:10.1103/PhysRevD.46.2141}}.

\bibitem{Yang:1993bp}
K.-C. Yang, W.~Y.~P. Hwang, E.~M. Henley, L.~S. Kisslinger, {QCD sum rules and
  neutron proton mass difference}, Phys. Rev. D 47 (1993) 3001--3012.
\newblock \href {https://doi.org/10.1103/PhysRevD.47.3001}
  {\path{doi:10.1103/PhysRevD.47.3001}}.

\bibitem{Belyaev:1985wza}
V.~M. Belyaev, B.~Y. Blok, {CHARMED BARYONS IN QUANTUM CHROMODYNAMICS}, Z.
  Phys. C 30 (1986) 151.
\newblock \href {https://doi.org/10.1007/BF01560689}
  {\path{doi:10.1007/BF01560689}}.

\bibitem{Antonov:2012ud}
D.~Antonov, J.~E. F.~T. Ribeiro, {Quark condensate for various heavy flavors},
  Eur. Phys. J. C 72 (2012) 2179.
\newblock \href {http://arxiv.org/abs/1209.0408} {\path{arXiv:1209.0408}},
  \href {https://doi.org/10.1140/epjc/s10052-012-2179-7}
  {\path{doi:10.1140/epjc/s10052-012-2179-7}}.

\bibitem{Li:2020rcg}
H.-D. Li, C.-D. L\"u, C.~Wang, Y.-M. Wang, Y.-B. Wei, {QCD calculations of
  radiative heavy meson decays with subleading power corrections}, JHEP 04
  (2020) 023.
\newblock \href {http://arxiv.org/abs/2002.03825} {\path{arXiv:2002.03825}},
  \href {https://doi.org/10.1007/JHEP04(2020)023}
  {\path{doi:10.1007/JHEP04(2020)023}}.

\bibitem{Ozdem:2022vip}
U.~\"Ozdem, {Electromagnetic properties of doubly heavy pentaquark states},
  Eur. Phys. J. Plus 137 (2022) 936.
\newblock \href {http://arxiv.org/abs/2201.00979} {\path{arXiv:2201.00979}},
  \href {https://doi.org/10.1140/epjp/s13360-022-03125-4}
  {\path{doi:10.1140/epjp/s13360-022-03125-4}}.

\bibitem{Ozdem:2022eds}
U.~\"Ozdem, {Electromagnetic form factors of the Bc-like tetraquarks: Molecular
  and diquark-antidiquark pictures}, Phys. Lett. B 838 (2023) 137750.
\newblock \href {http://arxiv.org/abs/2211.10169} {\path{arXiv:2211.10169}},
  \href {https://doi.org/10.1016/j.physletb.2023.137750}
  {\path{doi:10.1016/j.physletb.2023.137750}}.

\bibitem{Workman:2022ynf}
R.~L. Workman, et~al., {Review of Particle Physics}, PTEP 2022 (2022) 083C01.
\newblock \href {https://doi.org/10.1093/ptep/ptac097}
  {\path{doi:10.1093/ptep/ptac097}}.

\bibitem{Bose:1980vy}
S.~K. Bose, L.~P. Singh, {Magnetic Moments of Charmed and $B$ Flavored Hadrons
  in {MIT} Bag Model}, Phys. Rev. D 22 (1980) 773.
\newblock \href {https://doi.org/10.1103/PhysRevD.22.773}
  {\path{doi:10.1103/PhysRevD.22.773}}.

\bibitem{Simonis:2016pnh}
V.~\v{S}imonis, {Magnetic properties of ground-state mesons}, Eur. Phys. J. A
  52~(4) (2016) 90.
\newblock \href {http://arxiv.org/abs/1604.05894} {\path{arXiv:1604.05894}},
  \href {https://doi.org/10.1140/epja/i2016-16090-5}
  {\path{doi:10.1140/epja/i2016-16090-5}}.

\bibitem{Zakharov:1968fb}
V.~I. Zakharov, L.~A. Kondratyuk, L.~A. Ponomarev, {Bremsstrahlung and
  determination of electromagnetic parameters of particles}, Yad. Fiz. 8 (1968)
  783--792.

\bibitem{Pascalutsa:2004je}
V.~Pascalutsa, M.~Vanderhaeghen, {Magnetic moment of the Delta(1232)-resonance
  in chiral effective field theory}, Phys. Rev. Lett. 94 (2005) 102003.
\newblock \href {http://arxiv.org/abs/nucl-th/0412113}
  {\path{arXiv:nucl-th/0412113}}, \href
  {https://doi.org/10.1103/PhysRevLett.94.102003}
  {\path{doi:10.1103/PhysRevLett.94.102003}}.

\bibitem{Pascalutsa:2005vq}
V.~Pascalutsa, M.~Vanderhaeghen, {Chiral effective-field theory in the
  Delta(1232) region: I. Pion electroproduction on the nucleon}, Phys. Rev. D
  73 (2006) 034003.
\newblock \href {http://arxiv.org/abs/hep-ph/0512244}
  {\path{arXiv:hep-ph/0512244}}, \href
  {https://doi.org/10.1103/PhysRevD.73.034003}
  {\path{doi:10.1103/PhysRevD.73.034003}}.

\bibitem{Pascalutsa:2007wb}
V.~Pascalutsa, M.~Vanderhaeghen, {Chiral effective-field theory in the
  Delta(1232) region. II. Radiative pion photoproduction}, Phys. Rev. D 77
  (2008) 014027.
\newblock \href {http://arxiv.org/abs/0709.4583} {\path{arXiv:0709.4583}},
  \href {https://doi.org/10.1103/PhysRevD.77.014027}
  {\path{doi:10.1103/PhysRevD.77.014027}}.

\bibitem{Kotulla:2002cg}
M.~Kotulla, et~al., {The Reaction gamma p ---\ensuremath{>} pi zero gamma-prime
  p and the magnetic dipole moment of the delta+(1232) resonance}, Phys. Rev.
  Lett. 89 (2002) 272001.
\newblock \href {http://arxiv.org/abs/nucl-ex/0210040}
  {\path{arXiv:nucl-ex/0210040}}, \href
  {https://doi.org/10.1103/PhysRevLett.89.272001}
  {\path{doi:10.1103/PhysRevLett.89.272001}}.

\bibitem{Drechsel:2001qu}
D.~Drechsel, M.~Vanderhaeghen, {Magnetic dipole moment of the Delta+ (1232)
  from the gamma p ---\ensuremath{>} gamma pi0 pi reaction}, Phys. Rev. C 64
  (2001) 065202.
\newblock \href {http://arxiv.org/abs/hep-ph/0105060}
  {\path{arXiv:hep-ph/0105060}}, \href
  {https://doi.org/10.1103/PhysRevC.64.065202}
  {\path{doi:10.1103/PhysRevC.64.065202}}.

\bibitem{Machavariani:1999fr}
A.~I. Machavariani, A.~Faessler, A.~J. Buchmann, {Field-theoretical description
  of electromagnetic Delta resonance production and determination of the
  magnetic moment of the $\Delta^+(1232)$ resonance by the $e p \to e'N'\pi'
  \gamma'$ and $\gamma p \to N' \pi' \gamma'$ reactions}, Nucl. Phys. A 646
  (1999) 231--257, [Erratum: Nucl.Phys.A 686, 601--603 (2001)].
\newblock \href {https://doi.org/10.1016/S0375-9474(98)00611-3}
  {\path{doi:10.1016/S0375-9474(98)00611-3}}.

\bibitem{Drechsel:2000um}
D.~Drechsel, M.~Vanderhaeghen, M.~M. Giannini, E.~Santopinto, {Inelastic photon
  scattering and the magnetic moment of the Delta (1232) resonance}, Phys.
  Lett. B 484 (2000) 236--242.
\newblock \href {http://arxiv.org/abs/nucl-th/0003035}
  {\path{arXiv:nucl-th/0003035}}, \href
  {https://doi.org/10.1016/S0370-2693(00)00654-7}
  {\path{doi:10.1016/S0370-2693(00)00654-7}}.

\bibitem{Chiang:2004pw}
W.-T. Chiang, M.~Vanderhaeghen, S.~N. Yang, D.~Drechsel, {Unitary model for the
  gamma p ---\ensuremath{>} gamma pi0 p reaction and the magnetic dipole moment
  of the Delta+ (1232)}, Phys. Rev. C 71 (2005) 015204.
\newblock \href {http://arxiv.org/abs/hep-ph/0409078}
  {\path{arXiv:hep-ph/0409078}}, \href
  {https://doi.org/10.1103/PhysRevC.71.015204}
  {\path{doi:10.1103/PhysRevC.71.015204}}.

\bibitem{Machavariani:2005vn}
A.~I. Machavariani, A.~Faessler, {Propagator of the Delta resonance and
  determination of the magnetic moment of the Delta+ from the gamma p
  ---\ensuremath{>} gamma pi0 p reaction}, Phys. Rev. C 72 (2005) 024002.
\newblock \href {https://doi.org/10.1103/PhysRevC.72.024002}
  {\path{doi:10.1103/PhysRevC.72.024002}}.

\bibitem{LopezCastro:1997dg}
G.~Lopez~Castro, G.~Toledo~Sanchez, {Effects of the magnetic dipole moment of
  charged vector mesons in their radiative decay distribution}, Phys. Rev. D 56
  (1997) 4408--4411.
\newblock \href {http://arxiv.org/abs/hep-ph/9707202}
  {\path{arXiv:hep-ph/9707202}}, \href
  {https://doi.org/10.1103/PhysRevD.56.4408}
  {\path{doi:10.1103/PhysRevD.56.4408}}.

\bibitem{GarciaGudino:2013alv}
D.~Garc\'\i{}a Gudi\~no, G.~Toledo~S\'anchez, {Determination of the magnetic
  dipole moment of the rho meson using 4 pion electroproduction data}, Int. J.
  Mod. Phys. Conf. Ser. 35 (2014) 1460463.
\newblock \href {http://arxiv.org/abs/1305.6345} {\path{arXiv:1305.6345}},
  \href {https://doi.org/10.1142/S2010194514604633}
  {\path{doi:10.1142/S2010194514604633}}.

\end{thebibliography}
\bibliographystyle{elsarticle-num}

\end{document}